\RequirePackage{fix-cm}
\documentclass[smallextended]{svjour3}       % onecolumn (second format)
\smartqed  % flush right qed marks, e.g. at end of proof
\usepackage{graphicx}
\usepackage{url}
\usepackage{amsfonts}
\usepackage{verbatim}
\usepackage{footnote}
\usepackage{siunitx}
\usepackage{array}
\makesavenoteenv{tabular}
\makesavenoteenv{table}

\begin{document}

\title{A-Evac: the evacuation simulator for stochastic environment }
\subtitle{PRE-PRINT}

\author{Adam Krasuski \and Karol Krenski}

\institute{A. Krasuski \at
              The Main School of Fire Service \\
			  Slowackiego 52/54\\
              01-629 Warsaw, Poland\\
              \email{krasuski@inf.sgsp.edu.pl}           %  \\
			  \and
           K. Krenski \at
              The Main School of Fire Service \\
			  Slowackiego 52/54\\
              01-629 Warsaw, Poland\\
              \email{krenski@inf.sgsp.edu.pl}           %  \\
}

\date{Received: 24.11.2017 / Accepted: date}

\maketitle
%}}}
\begin{abstract}%{{{
We introduce an open-source software Aamks for fire risk assessment.
This article focuses on a component of Aamks -- an evacuation simulator
named a-evac. A-evac models evacuation of humans in the fire environment
produced by CFAST fire simulator. In the article we discuss the
probabilistic evacuation approach, automatic planning of exit routes,
the interactions amongst the moving evacuees and the impact of smoke on
the humans. The results consist of risk values based on FED, F-N curves
and evacuation animations.

\end{abstract}
%}}}

\section{Introduction}\label{intro}%{{{

\begin{quote}
\emph{
These days, there's not much you can understand about what is
going on around you if you do not understand the uncertainty
attached to pretty much every phenomenon.}

\hspace*\fill{\small--- J. N. Tsitsiklis}
\end{quote}

There is a continuous progress in understanding the fire phenomenon,
its impact on the structure and on the reaction of humans and safety
systems. There are scientific methods and models of the fire and the
emergency scene, there are computer implementations, but the complexity
of the domain impedes the more widespread use of these tools.

Currently, the most typical approach for assessing the safety of the
building is a precise choosing of the input parameters for a small
number of lengthy, detailed simulations. This procedure is managed by a practitioner,
based on his experience. However, based on \emph{heuristics and
biases}~\cite{tversky1974judgment,kahneman2009conditions} we have
concerns that human judgement surpasses statistical calculations. The
alternative is to let computer randomly choose the parameters and run
thousands of simulations. The resulting collection allows us, after
further processing, to judge on the safety of the building.

\section{Aamks, the multisimulations platform}\label{intro}%{{{

We created Aamks -- the platform for running simulations of fires and
then running the evacuation simulations, but thousands of them for a
single project. This is the Monte-Carlo approach. We use CFAST, which is
a rough, but a fast fire simulator. This allows us to explore the space
of possible scenarios and assess the probability of them. The second
component of risk -- consequences -- is taken from an evacuation
simulator capable to model evacuation in the fire environment. We use
a-evac as the evacuation simulator which we have built from scratch. The
\emph{multisimulation} is a handy name for what we are doing. Aamks
tries to assess the risk of failure of humans evacuation from a building
under fire. We applied methodology proposed
in~\cite{hostikka2002probabilistic,frame,PRA} -- stochastic simulations
based on the Simple Monte-Carlo approach~\cite{christian2007monte}. Our
primary goal was to develop an easy to use engineering tool rather than
a scientific tool, which resulted in: AutoCAD plugin for creating
geometries, web based interface, predefined setups of materials and
distributions of various aspects of buildings features, etc. The
workflow is as follows: The user draws a building layout or exports an
existing one. Next, the user defines a few parameters including the type
of the building, the safety systems in the building, etc. Finally, they
launch a defined number of stochastic simulations. As a result they
obtain the distributions of the safety parameters, namely: available
safe egress time (ASET), required safe egress time (RSET), fractional
effective dose (FED), hot layer height and temperature and F-N curves as
well as the event tree and risk matrix. 

Fortran is a popular language for coding simulations of physical
systems. CFAST and FDS+Evac are coded in fortran. Since we don't create
a fire simulator we code Aamks in python which is more comfortable
due to extremely rich collection of libraries. We decided that
borrowing Evac from FDS and integrating it with Aamks would be harder
for us than to code our own evacuation simulator, hence a-evac was born.
There's also a higher chance of attracting new python developers than
fortran developers for our project.

Aamks consists of the following modules: 

\begin{itemize}
\item{a-geom, geometry processing: AutoCAD plugin, importing
geometry, extracting topology of the building, navigating in the building,
etc. }
\item{a-evac, directing evacuees across the building and altering
their states } 
\item{a-fire, CFAST and FDS binaries and processing of their outputs } 
\item{a-gui, web application for user's input and for the results
visualisation }
\item{a-montecarlo, stochastic producer of thousands of input files for
CFAST and a-evac }
\item a-results, post-processing the results and creating the content
for reports, 
\item{a-manager, managing computations on the grid/cluster of computers }
\item{a-installer}
\end{itemize}

%}}}
\section {A-evac, the evacuation simulator}%{{{
In the following subsections we describe the internals of a-evac,
sometimes with the necessary Aamks context.
%}}}
\subsection{Geometry of the environment}\label{geom}%{{{

The Aamks workflow starts with a 3D geometry where fires and
evacuations will be simulated. We need to represent the building, which
contains one or more floors. Each floor can consist of compartments and
openings in them, named respectively COMPAS and VENTS in CFAST. Our
considerations are narrowed to rectangular geometries. There are two
basic ways for representing architecture geometries: a) cuboids can
define the insides of the rooms (type-a-geometry) or b) cuboids can
define the walls / obstacles (type-b-geometry). CFAST uses the
type-a-geometry. We create CFAST geometries from the input files of the
following format (there are more entities than presented here):

\begin{verbatim}
{
"FLOOR 1":
    {
      "ROOM": [
        [ [ 3.0 , 4.8 , 0.0 ] , [ 4.8 , 6.5 , 3.0 ] ] ,
        [ [ 3.0 , 6.5 , 0.0 ] , [ 6.8 , 7.4 , 3.0 ] ] 
      ] ,
      "COR": [
        [ [ 6.2 , 0.2 , 0.0 ] , [ 7.6 , 4.8 , 3.0 ] ] 
      ] ,
      "D": [
        [ [ 3.9 , 3.4 , 0.0 ] , [ 4.8 , 3.4 , 2.0 ] ] 
      ] ,
      "W": [
        [ [ 1.2 , 3.4 , 1.0 ] , [ 2.2 , 3.4 , 2.0 ] ] 
      ] ,
      "HOLE": [
        [ [ 3.0 , 6.5 , 0.0 ] , [ 4.8 , 6.5 , 3.0 ] ] 
      ] 
    }
}
\end{verbatim}

\begin{figure}[htp]
\begin{center}
\includegraphics[width=0.5\textwidth]{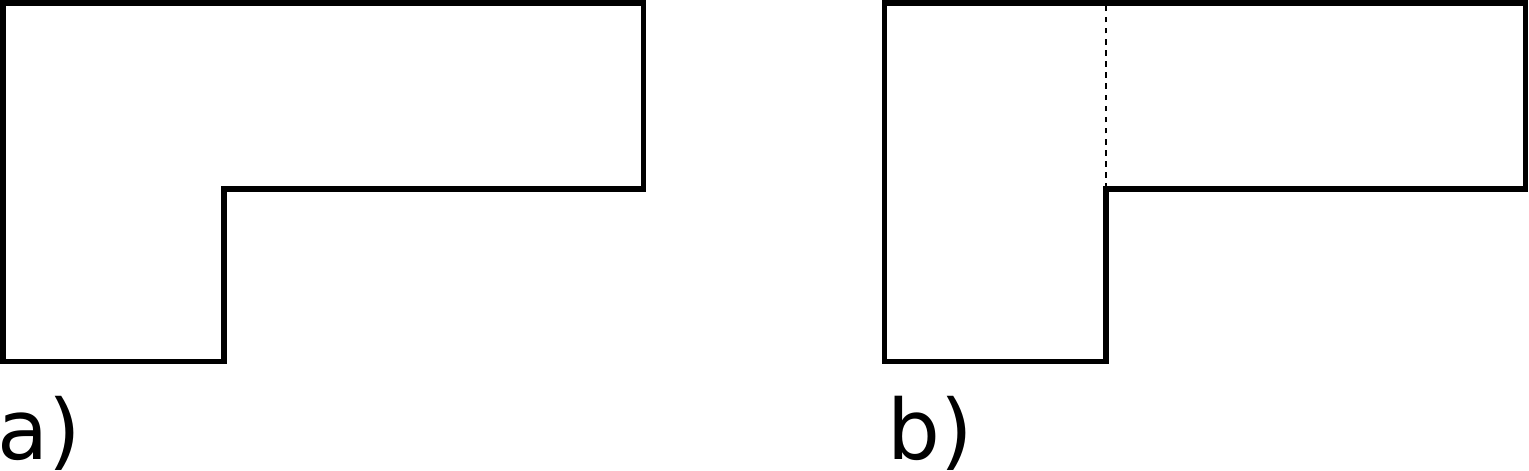} 
\end{center}
\caption{The concept of a HOLE: a) the room in reality, b) the room
representation in CFAST: two rectangles for separated calculations, but
open to each other via HOLE.}
\label{hole} 
\end{figure}

ROOM and COR(RIDOR) belong to COMPAS. D(OOR), W(INDOW) and HOLE belong
to VENTS. HOLE is a result of CFAST restrictions -- it is an artificial
entity which serves to merge two compartments into a single compartment
as shown on Figure~\ref{hole}.

All the entities in the example belong to the same FLOOR 1. The
triplets are $(x_0,y_0,z_0)$ and $(x_1,y_1,z_1)$ encoding the beginning
and the end of each entity in 3D space. In practice we obtain these
input files from AutoCAD, thanks to our plugin which extracts data from
AutoCAD drawing. There's also an Inkscape svg importer -- useful, but
without some features. Adding basic support for another graphics tools
is not much work.

\begin{figure}[htp]
\begin{center}
\includegraphics[width=0.9\textwidth]{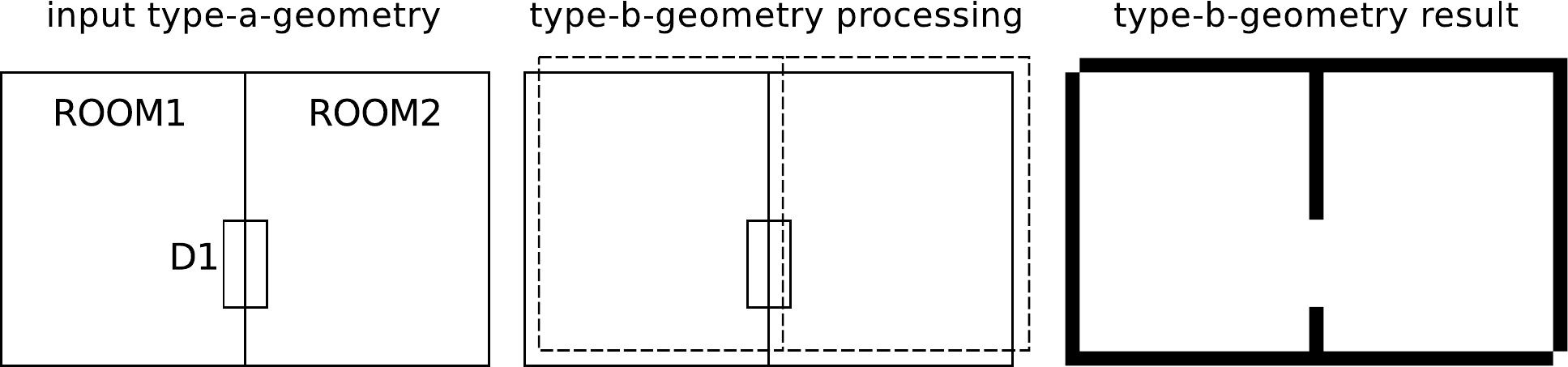} 
\end{center}
\caption{The conversion from type-a-geometry to type-b-geometry}
\label{toObstacles} 
\end{figure}

In later sections we will introduce problems of guiding evacuees
throughout the building. Those modules require the type-b-geometry. We
convert from a type-a-geometry to a type-b-geometry by duplicating the
geometry, translating the geometry and applying some logical operations.
Figure~\ref{toObstacles} shows the idea.

There are three aspects of movement when it comes to evacuation
modeling~\cite{cuesta2015evacuation}: (a) path-finding -- 
for the rough route out of the building, (b) local movement -- evacuees
interactions with other evacuees, with obstacles and with environment, and (c)
locomotion -- for "internal" movement of the agent (e.g. body sway).
A-evac models only (a) and (b).

%}}}
\subsection{Path-finding (roadmap)}%{{{

The simulated evacuees need to be guided out of the building.
The type-b-geometry provides the input for path-finding. Each
of the cuboids in type-b-geometry -- representing obstacles --  is
defined by the coordinates. These coordinates represent corners of the
shapes. Since we model each of the floors of a building separately, we
flatten 3D geometry into 2D and represent obstacles as rectangles.
Therefore type-b-geometry in path-finding module is represented as set
of 4-tuple coordinates $\big((x_0, y_0),(x_1, y_1),(x_2, y_2),(x_3,
y_3)\big)$. 

\begin{figure}[htp]
\begin{center}
\includegraphics[width=1.0\textwidth]{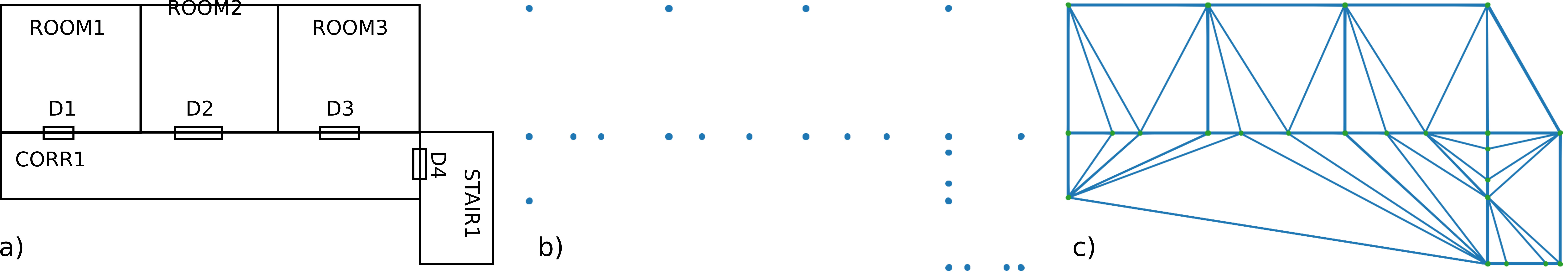} 
\end{center}
\caption{The idea of triangulation. a) original geometry, b)
bag-of-coordinates, c) triangulation.}
\label{tri} 
\end{figure}

The set of 4-tuple elements is then flatten to the set of coordinates --
bag-of-coordinates. Due to the fact that majority of the obstacles share
the coordinates, we remove duplicates from the set (for the sake of
performance). Then, this bag-of-coordinates is the input for
triangulation. We apply Delaunay
triangulation~\cite{delaunay1934sphere}, that represents space as a set
of triangles. Figure~\ref{tri} depicts the idea of triangulation. 

The triangles are used as navigation meshes for the agents. The
navigation meshes define which areas of an environment are traversable
by agents. 

After triangulation of bag-of-coordinates, some of the triangles are
located inside the obstacles -- those (by definition) are not
traversable so we remove them. What is left is a traversable-friendly
space.

We create then the graph of spatial transitions for the agents, based on
the adjacency of triangles obtained from the triangulation. Spatial
transition means that an agent can move from one triangle to another. 

An agent on an edge of a triangle can always reach the other two edges.
For triangles which share edges it allows an agent to travel from one
triangle to another.

The pairs of all neighbouring edges are collected. We use python
networkx module~\cite{brandes2005network} which creates a graph made of
the above pairs. For further processing we add agents positions to the
graph, by pairing them with the neighbouring edges. 

\begin{figure}[htp]
\begin{center}
\includegraphics[width=0.99\textwidth]{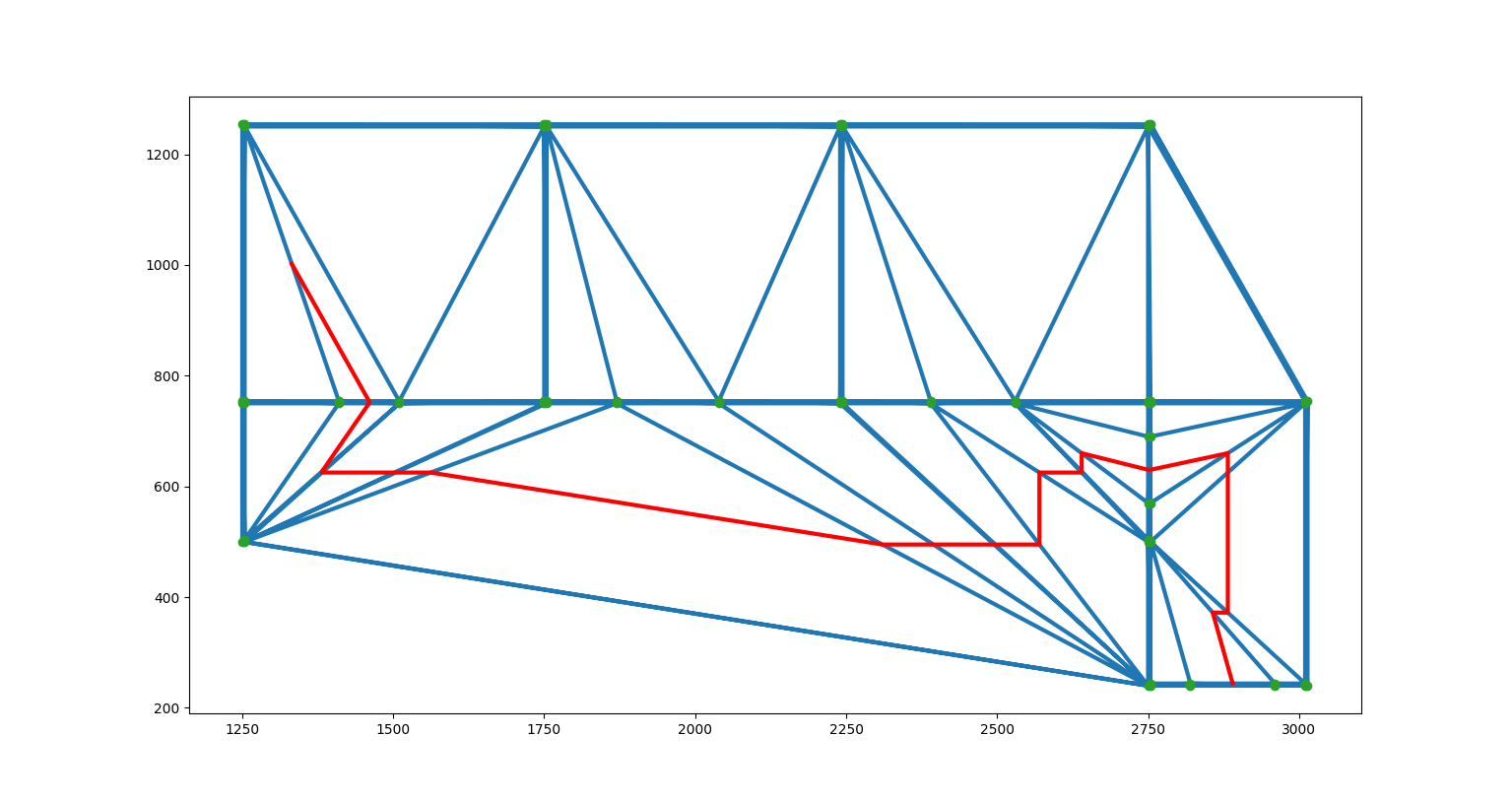} 
\end{center}
\caption{The roadmap defined by the graph for an example query. The red line
crosses centers of edges that an agent needs to travel to reach the exit.}
\label{evacgraph} 
\end{figure}

The graph represents all possible routes from any node to any other node
in the graph. We can query the graph for the route from the current
agent's position to the closest exit. It means that agent will walk
through the consecutive nodes and will finally reach the exit door. We
instruct networkx that we need the shortest distances in our routes
(default is the least hops on the graph) and we obtain a set of edges
the agent should traverse in order to reach the exit.
Figure~\ref{evacgraph} depicts the set of edges returned by the graph
for an example query.

The set of edges returned by the graph cannot be used directly for
path-finding. Neither the vertices of the edges nor the centers of them,
do define the optimal path that would be naturally chosen by evacuees during real
evacuation. Therefore an extra algorithm should be used to smooth the
path. For this purpose we apply funnel algorithm defined
in~\cite{chazelle1982theorem}. The funnel is a simple algorithm finding
straight lines along the edges. 

The input for the funnel consists of a set of ordered edges (named
portals) from the agent origin to the destination. The funnel always
consist of 3 entities: the origin (apex) and the two vectors from apex to 
vertices on edges -- the left leg and the right leg. 

The apex is first set to the origin of the agent and the legs are set to
the vertices of the first edge. We advance the left and right legs to
the consecutive edges in the set and observe the angle between the legs.
When the angle gets smaller we accept the new vertex for the leg.
Otherwise the leg stays at the given vertex. After some iteration one of
the legs should cross the other leg defining the new position of the
apex. The apex is moved and we restart the procedure.

\begin{figure}[htp]
\begin{center}
\includegraphics[width=0.59\textwidth]{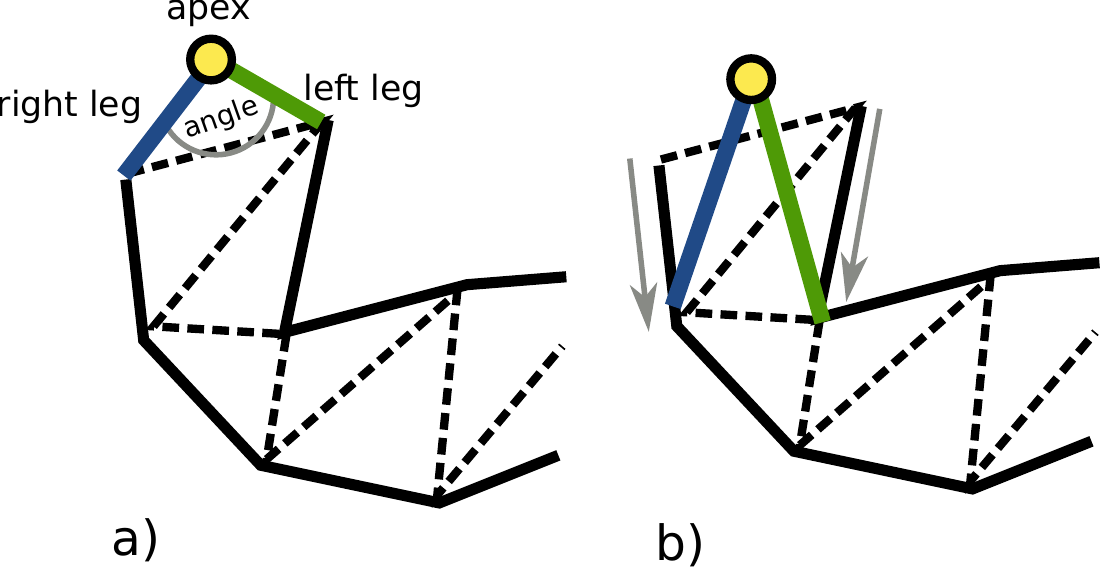} 
\end{center}
\caption{The idea of funnel algorithm. a) starting point, b) advancing
legs.}
\label{funnel} 
\end{figure}

As a result the path is smoothened and defined only by the points where the
changes in velocity vector are needed. Moreover, we used an improved version
of the funnel algorithm that allows for defining points keeping a
distance from the corners reflecting the size of the evacuee. This
allows for modeling the impaired evacuees on wheeled chairs or beds in
the hospitals. Figure~\ref{funnel} depicts the smoothened path by funnel
algorithm. 

\begin{figure}[htp]
\begin{center}
\includegraphics[width=0.99\textwidth]{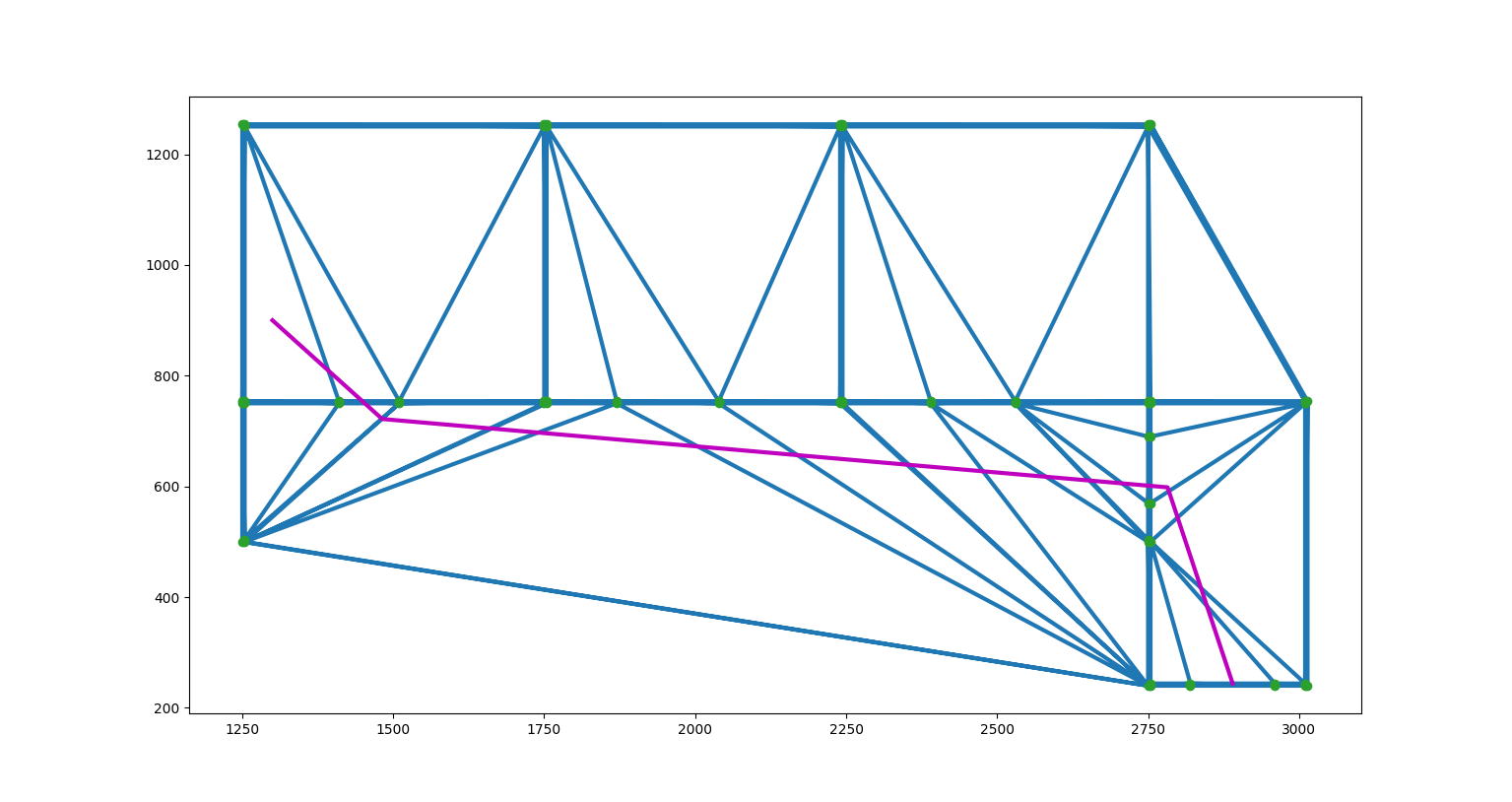} 
\end{center}
\caption{The roadmap from starting point to exit smoothened by
funnel algorithm. }
\label{funnel} 
\end{figure}

%}}}
\subsection{Local movement}\label{orca}%{{{

Local movement focuses on the interaction with (a) other agents (b) static
obstacles (walls) and (c) environmental conditions. A-evac handles (a) and
(b) via RVO2\footnote{\url{http://gamma.cs.unc.edu/RVO2/}} which is an
implementation of the Optimal Reciprocal Collision Avoidance (ORCA) algorithm
proposed in~\cite{van2008reciprocal,van2011reciprocal}. Later in this
section we describe how we are picking the local targets which is an
aspect of (b). (c) is basically altering agent's state such as speed.

RVO2 aims at avoiding the \emph{velocity
obstacle}~\cite{fiorini1998motion}. The velocity obstacle is the set of
all velocities of an agent that will result in a collision with another
agent or an obstacle. Otherwise, the velocity is collision-avoiding.
RVO2 aims at asserting that none of the agents collides with other
agents within time $\tau$. 

\begin{figure}[htp]
\begin{center}
\includegraphics[width=.7\textwidth]{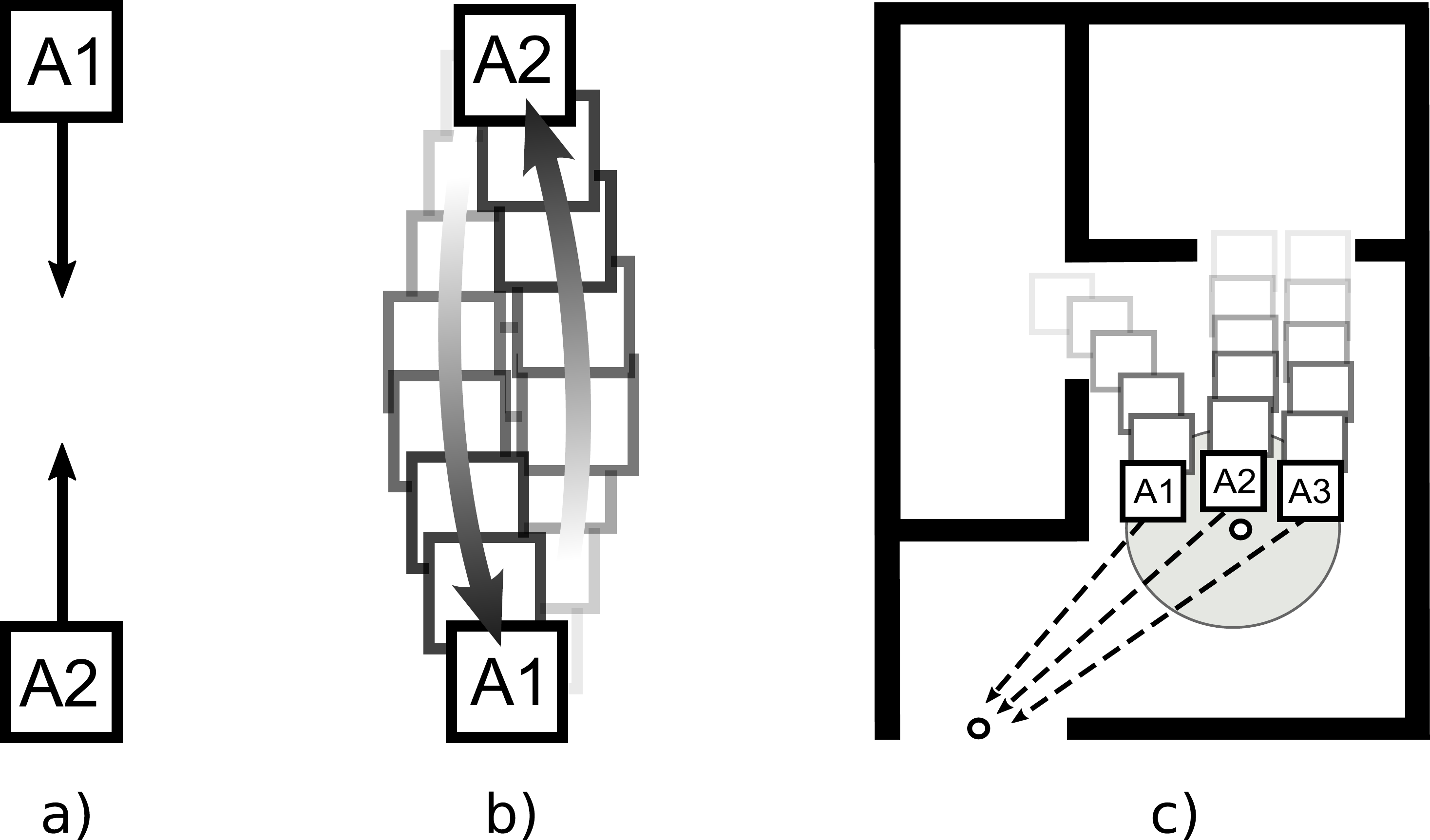} 
\end{center}
\caption{ RVO2 at its work of resolving collisions: (a) agents on direct
collision courses and (b) their calculated collision-avoiding courses, (c)
three agents crossing a HOLE in parallel.} 
\label{local_issues} 
\end{figure}

The overall approach is as follows: each of the agents is aware of other
agents parameters: their position, velocity and radius (agent's
observable universe). Besides, the agents have their private parameters:
maximum speed and a preferred velocity which they can auto-adjust
granted there is no other agent or an obstacle colliding. With each loop
iteration, each agent responses to what he finds in his surroundings,
i.e. his own and other agents radiuses, positions and velocities. The
agent updates his velocity if it is the velocity obstacle with another
agent. For each pair of colliding agents the set of collision-avoiding
velocities is calculated. RVO2 finds the smallest change required to
avert collision within time $\tau$ and that is how an agent gets his new
velocity. The agent alters up to half of his velocity while the other
colliding agent is required to take care of his half.
Figure~\ref{local_issues} a-b depicts the idea of velocity collision
avoidance. 

The algorithm remains the same for avoiding static obstacles. However,
the value of $\tau$ is smaller with respect to obstacles as agents
should be more 'brave' to move towards an obstacle if this is necessary
to avoid other agents. 

\begin{figure}[htp]
\begin{center}
\includegraphics[width=0.9\textwidth]{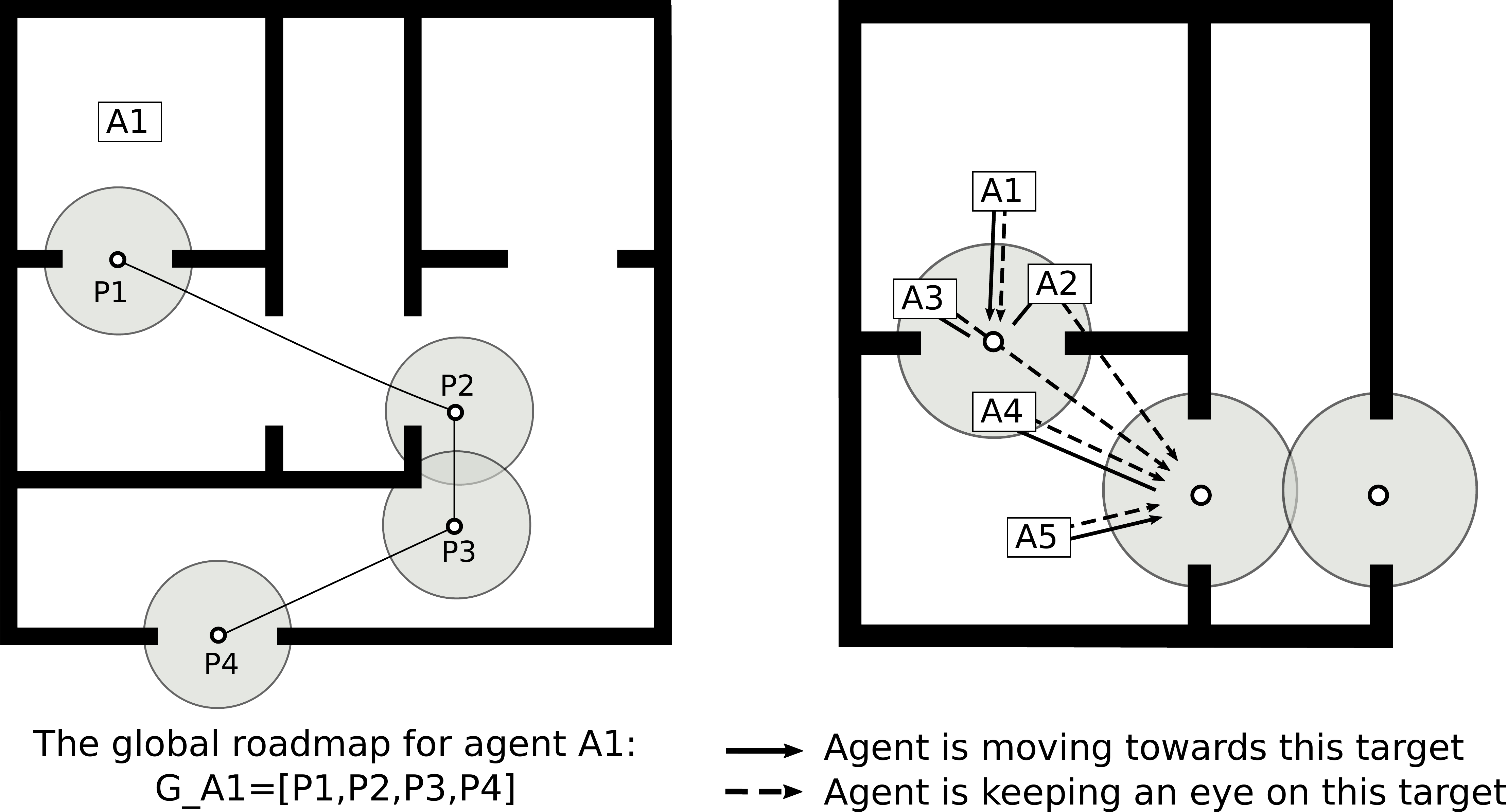} 
\end{center}
\caption{The roadmap and local movement} 
\label{path_vs_local} 
\end{figure}

It turned out problematic how to pick the local target from the
roadmap. Local targets need to be updated (usually advanced, but
not always) near points defined by funnel algorithm during path-fining
phase -- the \emph{disks} on Figure~\ref{path_vs_local}, after they become
visible to the agent. However, the disks can be crowded and agents can
be driven away from the correct courses by other agents. We carefully
inspected all possible states that agents can find themselves in. In
order to have a clearer insight and control over the agents inside the
disks we use the Finite State
Machine\footnote{\url{https://en.wikipedia.org/wiki/Finite-state_machine}}
instead of just plain algorithm block in our code. The state of the
agent is defined by 4 binary features: (a) is agent inside the disk? (b)
are \emph{where agent is walking to} and \emph{what agent is looking at}
the same target? (c) can agent see what he is looking at (or are there
obstacles in-between)? (d) has agent reached the final node? 

Within each iteration of the main loop we check the states of the
agents. The states can be changed by agents themselves -- e.g. agent
has crossed the border of the disk, or by our commands -- e.g. agent is
ordered to walk to another target. Consider these circumstances: the
agent has managed to see his next target and now he walks towards this
next target -- he is in state S1. But now he loses the eye contact with
this new target and finds himself in state S2. The program logic reacts
to such a state by transiting to the state S3: start looking at the previous
target and walk towards this previous target. Based on what happens next
we can order the transition to another state or just wait for the agent
to change the state himself. By careful examination of all possible
circumstances we can make sure that our states and their transitions can
handle all possible scenarios.

On Figure~\ref{local_issues}c) we show how agents are passing through a
HOLE. Due to our concept of the disks (where searching for new targets
takes place) and due to the internals of RVO2 we gain the desired effect
of agents not crossing the very center of the disk. Instead, the agents
can walk in parallel and advance to another target which looks natural
and doesn't create an unnecessary queue of agents eager to cross the
very center of the HOLE.

%}}}
\subsection{Evacuation under fire and smoke}%{{{

Each a-evac simulation is preceded with the simulation of the fire. We
have only tested a-evac with
CFAST~\cite{peacock2013CFAST,jones2000technical}. CFAST writes its
output to csv files. We need to query these CFAST results quite a bit,
therefore we transform and store these results in a fast in-memory
relational database\footnote{\url{https://www.sqlite.org/}}. For each
frame of time we are repeatedly asking the same questions: (a) given the
agent's coordinates, which room is he in? (b) what are the current
conditions in this room? 

When it comes to (b), the environment effects on the agent can be: (b.1)
limited visibility (eyes), (b.2) poisonous gases (nose) and (b.3)
temperature in the room (body). Both (b.1) and (b.2) are read from the
default (but configurable) height of 1.8~m. There are always two zones
in CFAST, which are separated at a known height, so we need to read the
conditions from the correct zone, based on where our 1.8~m belongs.

The value of visibility ($OD$ -- optical density) affects agent's speed.
We use the relation proposed in~\cite{frantzich2003utrymning} following
the FDS+Evac~\cite{korhonen2007fds}: 

\begin{equation}\label{speedF}
v_n^{pref} (K_s) = max \Big\{ v_{n, min}, v_n^{pref} \big(1 + \frac{\beta}{\alpha} \cdot K_s\ \big) \Big\}
\end{equation}

where: $K_s$ is the extinction coefficient ($[K_s] = m^{-1}$) calculated
as $OD/log_{10}e$ according
to~\cite{jin1974visibility,jin2016visibility}, $v_{n, min}$ is the
minimum speed of the agent $A_n$ and equals
$0.1\cdot~v_n^{pref}$(agent's preferable velocity), and $\alpha$,
$\beta$ are the coefficients defined in~\cite{frantzich2003utrymning}.  

Setting the minimal value of speed means that the agent does not stop
in thick smoke. They continue moving until the value of incapacitated
Fractional Effective Dose (FED) is exceeded, which is fatal to the
agent. FED is calculated from CFAST-provided amounts of the
following species in the agent environment: carbon monoxide (CO),
hydrogen cyanide (HCN), hydrogen chloride (HCl), carbon dioxide ($CO_2$)
and oxygen ($O_2$) by the equation~\cite{purser2002toxicity,korhonen2007fds}:

\begin{equation}
FED_{total} = (FED_{CO} + FED_{HCN} + FED_{HCl}) \times  HV_{CO_2} + FED_{O_2}
\end{equation}

where $HV_{CO_2}$ is the hyperventilation induced by the concentration
of $CO_2$. Following are the formulas for the terms in the above
equation. FEDs are given in ppm and time t in minutes. $C$ stands for
concentration of the species in \%:

\begin{equation}
FED_{CO} = \int_{0}^{t} 2.764 \times 10^{-5}(C_{CO}(t))^{1.036}dt
\end{equation}

\begin{equation}
FED_{HCN} = \int_{0}^{t} \frac{exp \big(\frac{C_{HCN}(t)}{43}\big)}{220} - 0.0045 dt
\end{equation}
Based on~\cite{hull2008hydrogen}

In contrast to the model applied in Evac, CFAST does not allow for
proactive correction of effect of nitrogen dioxide -- $C_{CN} = C_{HCN}
- C_{NO_2}$. Therefore this effect is not included in the calculations.

\begin{equation}
FED_{HCl} = \int_{0}^{t} \frac{C_{HCl}(t)}{1900} dt
\end{equation}
Based on~\cite{speitel1995toxicity,hull2008hydrogen}

\begin{equation}
FED_{O_2} = \int_{0}^{t} \frac{dt}{60 \cdot exp\big[8.13 - 0.54(20.9 - C_{O_2}(t))\big]}
\end{equation}

\begin{equation}
HV_{CO_2} = \frac{exp\big(0.1903 \cdot C_{CO_2} (t) + 2.0004\big)}{7.1}
\end{equation}

There are few quantitative data from controlled experiments concerning
the sublethal effect of the smoke on people. In
works~\cite{speitel1995toxicity,gann2008combustion,purser2002toxicity,gann2001sublethal,stec2010fire}
sublethal effect in a form of incapacitation ($IC_{50}$), escape ability
($EC_{50}$), \emph{lingering health problems} and \emph{minor effects}
were reported. Incapacitation was inferred from lethality data, to be
about one-third to one-half of those required for lethality. The mean
value of the ratios of the $IC_{50}$ to the $LC_{50}$ was 0.50 and the
standard deviation 0.21, respectively. In~\cite{gann2001sublethal} a
scale for effects based on FED was introduced. The three ranges were
proposed: 1 FED indicating lethality, 0.3 FED indicating incapacitation
and 0.01 FED indicating no significant sublethal effects should occur.
We propose based on this data a scale for sublethal effects of smoke for
evacuees as presented in Table~\ref{fed}. 

$FED_{total}$ affects the agent's movement in the smoke. For $FED_{total}>0.3$,
the smoke inhalation leads to sublethal
effects~\cite{gann2008combustion} -- the agent is not able to find
safety from the fire and just stays where he is. For $FED_{total} > 1$
we model lethal effects. We later use these effects in the final risk
assessment.

Table~\ref{fed} summarizes the FED effects on human health, what is our
original proposition for evaluation of sublethal effect of smoke. These
ranges are incorporeted in Aamks. It is based on the following
works:~\cite{speitel1995toxicity,gann2008combustion,purser2002toxicity,gann2001sublethal,stec2010fire}

\begin{table}[!h]\caption{FED effects on human health in Aamks.}\label{fed}
\begin{center}
\begin{tabular}{|c|c|}
  \hline
\textbf{FED} & \textbf{Effect on human health} \\
  \hline
$<0.01$ & Minor or negligible \\
  \hline
 $[0.01 - 0.3[$ & Low -- short period of hospitalization \\
  \hline
 $[0.3 - 1[$ & Heavy -- lingering health problems or permanent disability \\ 
  \hline
 $\ge 1 $ & Lethal \\ 
  \hline
\end{tabular}
\end{center}
\end{table} 

%}}}
\subsection{Probabilistic evacuation modeling}%{{{

This section presents the internals of our probabilistic evacuation
model, which we find distinct across the available, similar software. 

Table~\ref{stochastic_setup} presents the distributions of the input
parameters used in Aamks. Each of the thousands of simulations in a
single project is initialized with some random input setup according to
these distributions. Aamks has a library of the default parameters
values for important building categories (schools, offices, malls etc.).
The Aamks users should find it convenient to have all the distributions
in a library, but they may choose to alter these values, which is
possible. 

Most of the data in table~\ref{stochastic_setup} come from the standards
and from other models, mostly FDS/Evac. Following are some comments on
table~\ref{stochastic_setup}.

Aamks puts much attention to the pre-evacuation
time~\cite{cuesta2015evacuation}, which models how people lag before
evacuating after the alarm has sounded. Positions 7. and 8. are
separated, because the behaviour of humans in the room of fire origin is
distinct. We compile two regulations \emph{C/VM2 Verification Method:
Framework for Fire Safety Design}~\cite{nznorm2013} and \emph{British
Standard PD 7974-6:2004}~\cite{bs7974} in order to get the most
realistic, probability-based pre-evacuation in the room of fire origin
and in the rest of the rooms.

The Horizontal/Vertical speed (unimpeded, walking speed of an agent) is
based
on~\cite{hurley2015sfpe,fruin1971pedestrian,predtetschenski1978planning,helbing2002simulation}. 
 
Speed in the smoke is modeled by formula~\ref{speedF}. 

\begin{table}[!h]
\setlength\extrarowheight{3pt}
\caption{Parameters of the distributions for the exemplary scenario.}
\label{stochastic_setup}
\begin{center}
\begin{tabular}{ccccc}
    & Parameter                                  & Distribution & $\mu$/min & $\sigma$/max \\
  \hline
1.  & Denstity in rooms [$m^2/humans$]               & normal     & 5      & 2 \\
2.  & Densitiy on corridors [$m^2/humans$]           & normal     & 20     & 3 \\
3.  & Densitiy in stairways [$m^2/humans$]           & normal     & 50     & 3 \\
4.  & Human location in the compartment $x$          & uniform    & 0      & room width \\
5.  & Human location in the compartment $y$          & uniform    & 0	   & room depth \\
6.  & Time of the alarm                              & log-normal & 0.7    & 0.2 \\
7.  & Pre-evacuation time in the room of fire origin & uniform    & 0      & 30 \\
8.  & Pre-evacuation time in other compartments      & log-normal & 3.04   & 0.142 \\
9.  & Horizontal speed                               & normal     & 1.2    & 0.2 \\
10. & Vertical speed                                 & normal     & 0.7    & 0.2 \\
11. & $\alpha$ for speed in the smoke                & normal     & 0.706  & 0.069 \\
12. & $\beta$ for speed in the smoke                 & normal     & -0.057 & 0.015 \\
13. & Humans taking an alternative evacuation route  & binomial   & 0.03   & 0.97 \\
\end{tabular}
\end{center}
\end{table} 

The randomness of the simulations comes from the random number
generator's \emph{seed}. We save the seed for each simulation so that we
can repeat the very same simulation, which is useful for debugging and
visualisation.

We register all the random input setups and the corresponding results in
the database. We are expecting to research at some point the relationships
in these data with data mining or sensitivity analysis.

\begin{figure}[htp]
\begin{center}
\includegraphics[width=1\textwidth]{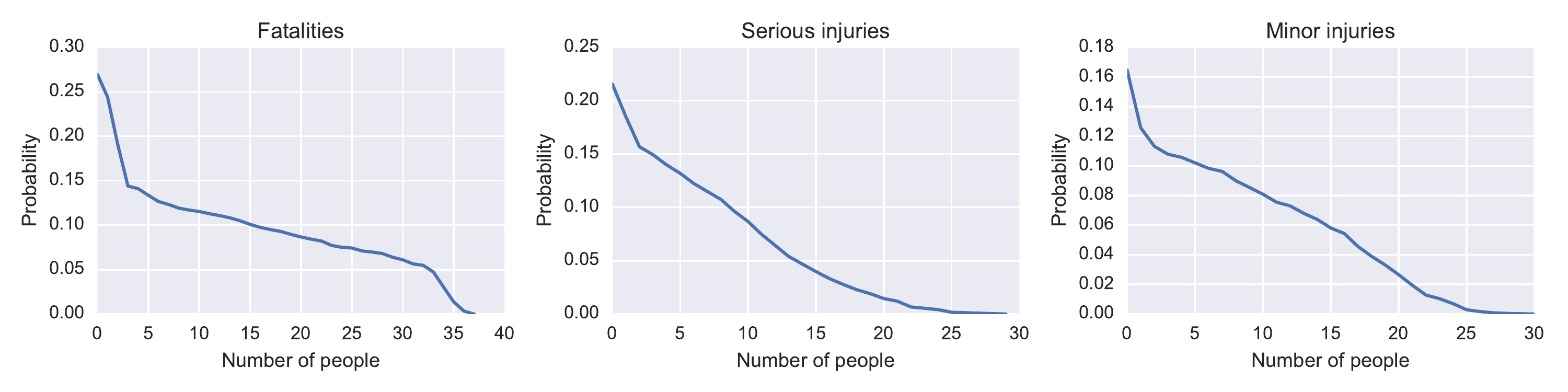} 
\end{center}
\caption{The results of evacuation modeling as F-N curves.} 
\label{ccdf} 
\end{figure}

The final result of Aamks is the compilation of multiple simulations as
a set of distributions i.e. F-N curves. The F-N curves were created as
in~\cite{frantzich2003utrymning}. Figure~\ref{ccdf} depicts the exemplary
results. 

%}}}
\subsection{Visualization}%{{{

\begin{figure}[htp]
\begin{center}
\includegraphics[width=.8\textwidth]{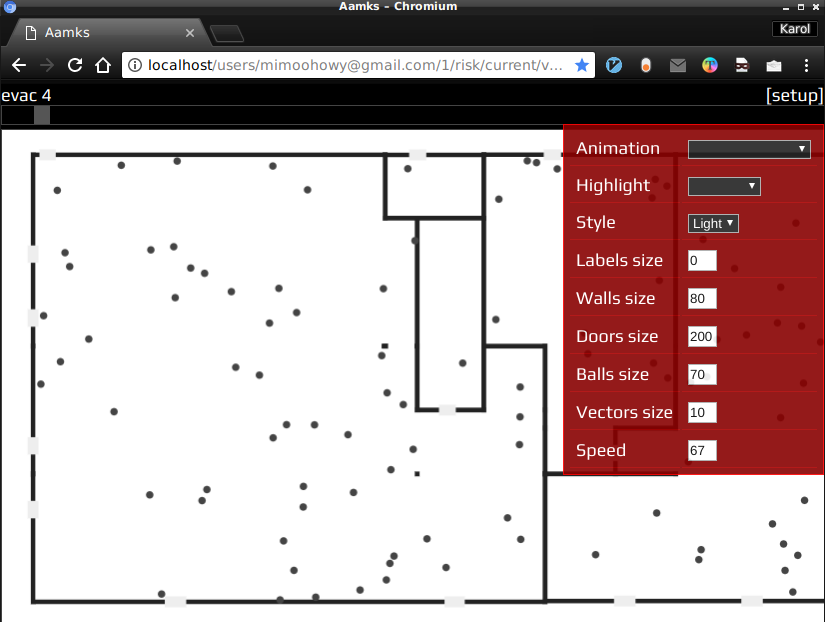} 
\end{center}
\caption{2D visualization: animation of evacuees } 
\label{2dvis} 
\end{figure}

In Aamks we use a 2D visualization for supervising the potential user's
faults in his CAD work (e.g. rooms with no doors (Figure~\ref{2dvis})), for
the final results, and for our internal developing needs. We use a web
based technology which allows for displaying both static images and the
animations of evacuees.

\begin{figure}[htp]
\begin{center}
\includegraphics[width=1\textwidth]{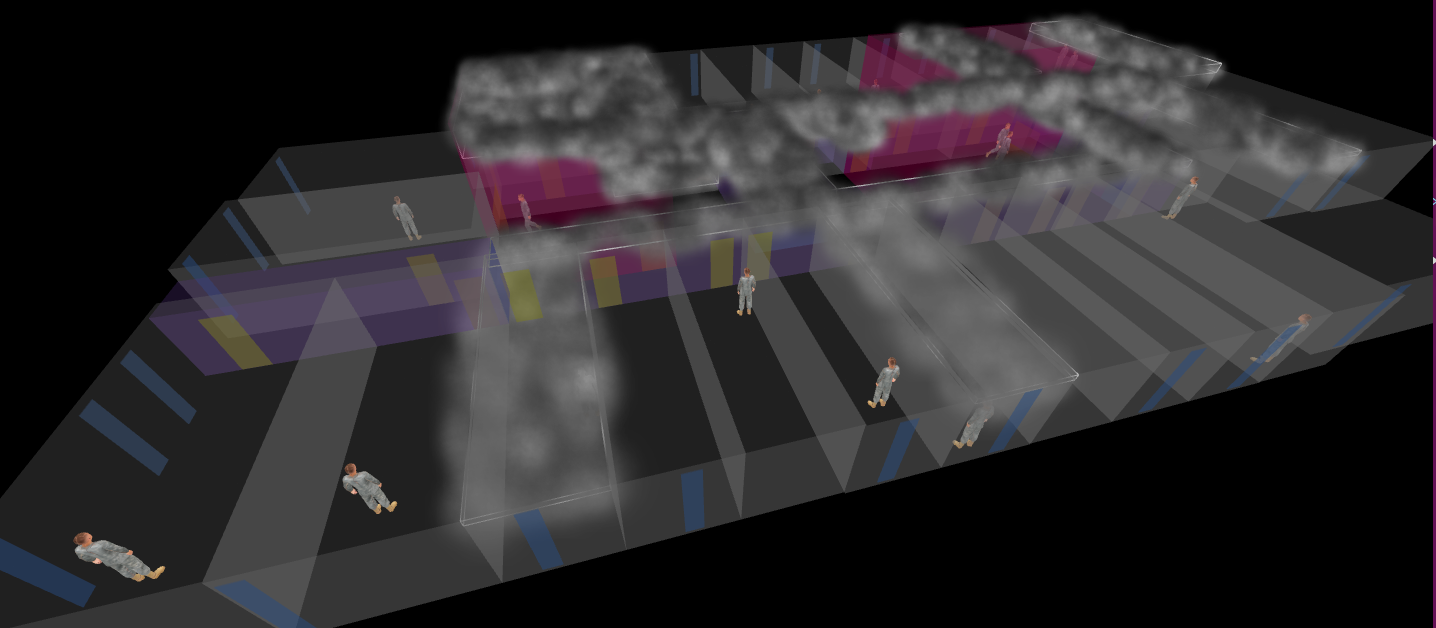} 
\end{center}
\caption{3D visualization} 
\label{3dvis} 
\end{figure}

We also have a web based 3D visualization made with WebGL
\texttt{Threejs}. This subsystem displays realistic animations of humans
during their evacuation under fire and smoke. (Figure~\ref{3dvis}).

%}}}

\section{Quality and the performance of a-evac}\label{qq}%{{{

Below we evaluate the quality of a-evac as described
in~\cite{cuesta2015evacuation,ronchi2013process} as well as it's
computer performance. 

%}}}
\subsection{Verification of a-evac}\label{vv}%{{{

Verification and validation deals with how close the results of the
simulations to the reality are. We took care to be compliant with the
general development recommendations~\cite{cuesta2015evacuation} by: (1)
obeying good programming practices, (2) verifying intermediate simulation
outputs, (3) comparing simulation outputs against the analytical results,
and (4) creating debugging animations. 

The are three types of errors that can be generated by our software: a)
error in deterministic modeling of single scenario, b) error of Monte
Carlo approximation, c) statistical error  -- disturbance. 

For the first type of error we applied the methods proposed
in~\cite{ronchi2013process}. The proposed tests are organized in five
core components: (1) pre-evacuation time, (2) movement and navigation,
(3) exit usage, (4) route availability, and (5) flow
conditions/constraints. For each category there are detailed tests for
the geometry, the scenario and the expected results. The results are in
table~\ref{IMO}.

\begin{table}[!h]\caption{The results of Aamks tests}\label{IMO}
\setlength\extrarowheight{2pt}
\begin{center}
\begin{tabular}{llll}
Id.      & Name of the tests                   & Test code  & Results\\
  \hline
1.       & Pre-evacuation time distributions   & Verif.1.1  & OK \\
2.       & Speed in a corridor                 & Verif.2.1  & OK \\
3.       & Speed on stairs                     & Verif.2.2  & OK \footnote{The method is not straightforward}\\
4.       & Movement around the corner          & Verif.2.3  & OK \\
5.       & Assigned occupant demographics      & Verif.2.4  & OK \\
6.       & Reduced visibility vs walking speed & Verif.2.5  & OK \\
7.       & Occupant incapatication             & Verif.2.6  & OK \\
8.       & Elevator usage                      & Verif.2.7  & --\\
9.       & Horizontal counter-flows (rooms)    & Verif.2.8  & OK \\
10.      & Group behaviours                    & Verif.2.9  & --\\
11.      & People with movement disabilities   & Verif.2.10 & -- \\
12.      & Exit route allocation               & Verif.3.1  & OK \\
13.      & Social influence                    & Verif.3.2  & --\\
14.      & Affiliation                         & Verif.3.3  & --\\
15.      & Dynamic availability of exits       & Verif.4.1  & OK \\
16.      & Congestion                          & Verif.5.1  & OK \\
17.      & Maximum flow rates                  & Verif.5.2  & OK \\
\end{tabular}
\end{center}
\end{table} 

RVO2, the core library of a-evac which drives the local movement, was
also evaluated in~\cite{viswanathan2014quantitative}. The conclusions
are that RVO2 is of the quality comparable with the lattice gas and
social force models. The social force model is commonly used in a number
of evacuation software.

The above is the evaluation of a single, deterministic simulation.
However, the final result is the compilation of the whole collection of
such single simulations -- this is how we get the big picture of the
safety of the inquired building. The picture is meant to present risk.
Probability of risk is calculated as a share of simulations resulted in
fatalities, in the total number of simulations. 

The accuracy of this evaluation depends on the method applied --
stochastic simulations. The error is proportional to the square root of
number simulations. Namely for the discrete Bernoulli probability
distributions used for example for evaluation of probability of scenario
with fatalities, the error is calculated as follows: 

\begin{equation}
\hat{\sigma}_n = 1.96\sqrt{\frac{\hat{p}_n(1-\hat{p}_n)}{n}}; 
\end{equation}

where: $\hat{p}_n$ is the probability of fatalities obtained as a number
of simulations resulted in fatalities to the total number of
simulations, $n$ is the number of simulations.  

We are aware of the third type of error which may be generated by the
application. The input for Aamks is a set of various probability
distributions what may occasionally generate unreal scenario. For
example an evacuee who moves very slowly on corridors and very fast on
stairs. In most cases these errors are related to the other parts of
Aamks i.e. probabilistic fire modeling. However, fire environment
impacts the evacuation. This error can be evaluated by comparison of
data generated by Aamks with real statistics. So far we do not have idea
how to tackle this problem efficiently. We consider to evaluate this
error by launching simulations for the building stock and check
whether we reconstruct historical data. This method is very laborious
and not justified at that moment, because our application still lacks
some models i.e. fire service intervention -- what has significant
impact on fire.

%}}}
\subsection{Performance of the Model}%{{{

The main loop of Aamks processes all agents in the time iteration.
Table~\ref{timing} summarizes how costly the specific calculations for a
single agent within a single time iteration are. The tests were
performed on computer with Intel Core i5-2500K CPU at 3.30 GHz with 8 GB
of RAM. 

\begin{table}[!h]\caption{The costs of a single loop iteration per agent}
\label{timing}
\setlength\extrarowheight{3pt}
\begin{center}
\begin{tabular}{llr}
Activity & Time & Total share\\
\hline

Position update & \SI{8.12e-6}{\second} & 3.97 \%\\
Velocity update & \SI{1.07e-5}{\second} & 5.26 \%\\
Speed update    & \SI{8.50e-5}{\second} & 41.63 \%\\
State update    & \SI{7.75e-8}{\second} & 0.03 \%\\
Goals update    & \SI{9.25e-8}{\second} & 0.04 \%\\
FED update      & \SI{1.01e-4}{\second} & 48.98 \%\\
Time update     & \SI{1.22e-7}{\second} & 0.06 \%\\

\end{tabular}
\end{center}
\end{table} 

The total time of a single step of the simulation for one agent is
\SI{2e-4}{\second} and it grows linearly with the number of agents. The
speed and FED calculations are most costly, because they both make
database queries against the fire conditions in the compartment. The
time step for a-evac iteration is 0.05 s. There is no significant
change in fire conditions within this time frame. Therefore for
performance optimization we update speed and FED every 20-th step of
the simulation. 

%}}}

\section{Discussion}\label{discussion}%{{{

Vertical evacuation is troublesome and not implemented. There is RVO2
3D, but it is for aviation where agents can pass above each other --
clearly not for our needs. Besides, we think things look actually better
in 2D. We like the idea that vertical evacuation can be still considered
2D, just rotated, and we plan to move in this direction.

A-evac does not model the social or group behaviours. However, it is
difficult to evaluate, how much the lack of such functionalities impacts
the resulting probability distributions. 

In the workflow we run a CFAST simulation first. Then a-evac simulation
runs on top of CFAST results. This sequential procedure has it's
drawbacks, e.g. we don't know how long the CFAST simulation should last
to produce enough data for a-evac, so we run "too much" CFAST for
safety. Also, evacuees cannot trigger any events such as opening the
door, etc. We considered a closer a-evac-CFAST integration. 

There seems to be lot's of space for improvement in Aamks. We work with
the practitioners and know the reality of fire engineering. We know the
limitations of our current implementations and most of them can be
addressed -- there are models and approaches that we can implement and
the major obstacle are the limits of our team resources. Therefore we
invite everyone interested to join our project at
\url{http://github.com/aamks}.

%}}}
\section{Conclusion}\label{con}%{{{

Aamks is actively developed since 2016 and we are truly engaged in
making it better. The software, though not really ready for end-users
has already served as a support for commercial projects and fire
engineers and scientists regard Aamsk as having potential. The
stochastic based workflow of Aamks is not a new concept. There are
opinions in the community that this approach is how fire engineering
should be done. Since no wide-used implementation has been created so
far, this is an additional motivation that drives our project.

% Aamks compiles the large collection of CFAST/a-evac simulations into the
% resulting output distributions, which are further analyzed in terms of risk
% assessment. 

%}}}
\begin{acknowledgements}%{{{

Aamks is hosted at github at: \url{http://gitub.com/aamks}.
This work was supported in part by ConsultRisk LTD and F\&K Consulting
Engineers LTD under research grants CR-2016 SiMo (Simulation Modules) . 

\end{acknowledgements}

%\bibliography{biblio}
%}}}
\end{document}